# Pseudo-random number generator based on asymptotic deterministic randomness

Kai Wang [1*] Wenjiang Pei[1], Haishan Xia[1], Yiu-ming Cheung[2]

(1.Department of Radio Engineering, Southeast University, Nanjing, China)

(2. Department of Computer Science, Hong Kong Baptist University, Hong Kong, China)

An approach to generate the pseudorandom-bit sequence from the asymptotic deterministic randomness system is proposed in this Letter. We study the characteristic of multi-value correspondence of the asymptotic deterministic randomness constructed by the piecewise linear map and the noninvertible nonlinearity transform, and then give the discretized systems in the finite digitized state space. The statistic characteristics of the asymptotic deterministic randomness are investigated numerically, such as stationary probability density function and random-like behavior. Furthermore, we analyze the dynamics of the symbolic sequence. Both theoretical and experimental results show that the symbolic sequence of the asymptotic deterministic randomness possesses very good cryptographic properties, which improve the security of chaos based PRBGs and increase the resistance against entropy attacks and symbolic dynamics attacks.

## I. INTRODUCTION

Pseudorandom bit generators (PRBGs), which can be easily implemented with simple and fast software routines, play a central role in modern security schemes, such as generating cryptographic keys and initializing variables in cryptographic protocols randomly. Though the sequences generated by PRBGs appear to be random, they are not ideal or truly random bit sequences, because they can be reproduced by certain deterministic algorithms on finite-state. When the periods of those pseudorandom bit sequences are very long, they can pass many statistical tests.

Chaotic systems are indeed characterized by ergodicity, sensitive dependence on initial conditions and random-like behaviors, properties which seem pretty much the same required by several primitives such as diffusion and confusion operations in conventional cryptography. Today, we have exploited chaotic systems for PRBGs in order to obtain some unpredictable behaviors from them. For example, Refs. [1-2] are devoted to the analysis of the application of a chaotic piecewise-linear map as PRBGs and mathematically analyze the information generation process of a class of piecewise linear 1D maps. Effective PRBGs are obtained by means of a chaotic system based on a pipeline analog-to-digital converter [3]. The discretized chaos based PRBGs of low complexity are analyzed to evaluate its suitability for the integrated implementation [4-6].

With researches of chaotic cryptology going more thorough, several cryptanalysis schemes have been proposed to analyze the security of the chaotic based PRBGs [7-8]. Several cryptanalysis, such as one-time pad attacks and entropy attacks, are proposed in Ref.[8]. It also gives a

---

[*] Corresponding author.

E-mail addresses: kaiwang@seu.edu.cn

cryptanalysis strategy by using symbolic dynamics. According to chaotic symbolic sequence, the attacker can calculate the initial value or control parameter utilized as the secret key, because there is a one-to-one correspondence between initial value and symbolic sequences of generating partitions [8]. With those attack strategies, many symmetric and asymmetric encryptions, which use PRBGs based on chaotic systems, have been proved to be insecure [9-10].

In 1997, J.A. Gonzalez discussed dynamics of the general function: $x_n = \sin^2(\pi\theta z^n)$. The author discovers that the sequence produced by this function is unpredictable in the short term and has the characteristic of multi-value correspondence when $z$ is a relative prime fraction number. This phenomenon is named as "the deterministic randomness", in order to differentiate it from chaos and randomness. Further study shows that we can construct generalized asymptotic deterministic randomness systems by utilizing the piecewise linear/nonlinear map and the noninvertible nonlinearity transform [11-16]. In Ref.[17], we have proved that the general function can generate deterministic randomness only when the value of parameter $z$ belongs to some relative prime fraction number which is larger than one. We have further proved that any realizable models for deterministic randomness will degenerate to some special high dimensional chaotic system. Furthermore, we analyze the underlying dynamics such as fixed point, bifurcation process, Lyapunov exponent spectrum, and symbolic dynamics etc. in detail.

Obviously, the asymptotic deterministic randomness can not only become a dominant approach in exploring the relationship between chaos and randomness, but also can be associated with some famous number theoretic concepts and open problems in number theory. In fact the asymptotic deterministic randomness systems can be a potential way of improving the security of the chaotic pseudorandom sequences, because the principle of designing a counter-assisted iterative number generator proposed by Shamir is similar to the asymptotic deterministic randomness systems [18]. In this paper, we will study the characteristic of multi-value correspondence of the asymptotic deterministic randomness, which is constructed by the piecewise linear map and the noninvertible nonlinearity transform; we also give the discretized systems in the finite digitized state space. The statistic characteristics, such as stationary probability density function and random-like behavior, are investigated numerically. Furthermore, we analyze the dynamics of the symbolic sequence. Both theoretical and experimental results show that the symbolic sequence of the asymptotic deterministic randomness possesses very good cryptographic properties, which improve the security of chaos based PRBGs and increase the resistance against entropy attacks and the symbolic dynamics attacks.

The rest of this paper is organized as follows. In Section 2, we focus on asymptotic deterministic randomness systems and the discretized systems. In Section 3, we discuss the statistic characteristics of asymptotic deterministic randomness sequences. In Section 4, we analyze the security of a stream cipher based on PRBGs of the asymptotic deterministic randomness. Finally, the concluding remarks are given in Section 5.

## II. ASYMPTOTIC DETERMINISTIC RANDOMNESS AND ITS DISCRETIZED SYSTEMS

Consider the equation $x_n = p(\theta T z^n)$, where $p(t)$ is the periodic function, $z$ is an integer, $\theta$ defines the initial condition, and $T$ is the period of function $p(t)$. As $p(t) = \sin^2(t)$ and $z = 2$, the equation is evidently the general solution of the Logistic map. Take the following function for instance: $x_n = \sin^2(\theta T z^n)$. Let $\theta = \theta_0 + q^m k$ and let $z$ be a rational number expressed as $z = p/q$, where $p, q$ are relative prime numbers, and $m, k$ are integers. We can see that, as given the sequence $x_0, x_1, x_2, \cdots x_m$ produced by $x_n = \sin^2(\theta T z^n)$, the next value $x_{m+1}$ can take $q$ possible values (this scenario is also called "multi-value correspondence"), resulting in the sequence unpredictable in the short term [11-14]. To be distinguished from chaos, this phenomenon is named as "deterministic randomness" [11-14]. In Ref.[17], we have constructed a series of generalized asymptotic deterministic randomness systems by utilizing the piecewise linear map and the non-invertible nonlinearity transform. Let's study the following two types of piecewise linear map:

$$h(a,t) = \mod(a*t, 1) \tag{1a}$$

$$g(a,t) = \begin{cases} \mod(a*t,1) & \mod(\lfloor t*a \rfloor, 2) = 0 \\ -\mod(a*t,1)+1 & \mod(\lfloor t*a \rfloor, 2) = 1 \end{cases} \tag{1b}$$

As shown in Eq.1a and Eq.1b, two objects $h(a,t)$ and $g(a,t)$ are considered homeomorphic, and are conjugate topological transform of $a$D-Chebyshev map when $a$ is a positive whole number. When $a = p/q > 2$, consider the following non-invertible nonlinearity transform:

$$x_{n+1} = h(a, x_n), \ y_n = h(b, x_n) \tag{2a}$$

$$x_{n+1} = g(a, x_n), \ y_n = g(b, x_n) \tag{2b}$$

**Theorem 1:** Consider the following nonlinear dynamical system $x_{n+1} = h(a, x_n), y_n = h(b, x_n)$ and $x_{n+1} = g(a, x_n), y_n = g(b, x_n)$, where $a = p/q > 2$ is a relative prime fraction number. When $b = q^N$, $y_n$ and $y_{n+m}$ ($m = 1, 2, \cdots N$) have the perfect multi-value correspondence with $p^m : q^m$ [17]. The first return maps are shown in Fig.1.

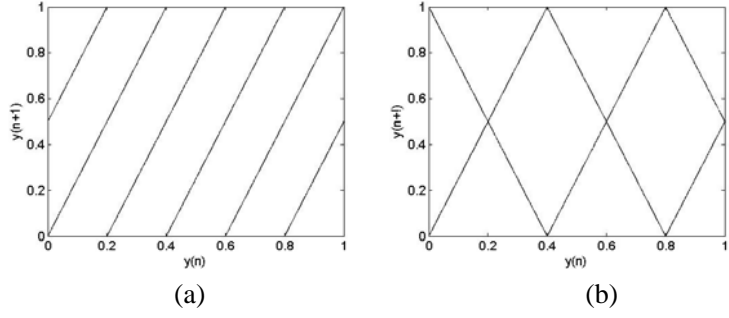

(a)                  (b)

FIG.1: First-return maps produced by Eq.2a and Eq.2b for $a = 5/2$: (a) $x_{n+1} = h(a, x_n)$, $y_n = h(b, x_n)$; (b) $x_{n+1} = g(a, x_n)$, $y_n = g(b, x_n)$

As shown in Eq.(2), the asymptotic deterministic randomness systems are constructed by the piecewise linear maps, which can be discretized in the finite digitized state space and implemented with low complexity digital hardware requirement[5-6]. Let $x_{n+1} = F(x) = ax_n \bmod 1$, its discretized model with the truncation approximation strategies is given below:

$$F_t(x') = [a \Box x']_r (\bmod 1) = \lfloor 2^n \Box a \Box x'(\bmod 2^n) \rfloor \Box 2^{-n}$$

where $x'$ is univocally determined by the binary representation of the $n$-bit natural number. Let $x' \Box 2^n = k$, $0 \le k < 2^n, k \in \mathbb{N}$, and then $f_r(k) = 2^n F_t(2^n x') = \lfloor a \Box k \rfloor (\bmod 2^n)$, where $f_r : [0, 2^n - 1] \to [0, 2^n - 1]$. Let $a = \lfloor a \rfloor + \beta$, and then the realizable discretized models with low complexity digital hardware requirement can be described as follows:

$$\lfloor ak \rfloor (\bmod m) = \lfloor ak \rfloor - \lfloor m^{-1} \lfloor ak \rfloor \rfloor m = \lfloor a \rfloor k + \lfloor \beta k \rfloor - \lfloor m^{-1}(\lfloor a \rfloor k + \lfloor \beta k \rfloor) \rfloor m \quad (3)$$

As shown in Eq.(3), the piecewise linear map $F : [0,1] \to [0,1]$ is discretized as the map with the finite digitized state space $f_r : [0, 2^n - 1] \to [0, 2^n - 1]$. Let $y_n = F'(x_n) = bx_n \bmod 1$, and then its discretized model in the finite digitized state space is: $f'(k) = 2^n F(2^n x') = b \Box k (\bmod 2^n)$. Let the positive integer sequences $\{X_i\}, \{Y_i\} \in \{0, 1, \cdots, 2^n - 1\}$, and then the discretized asymptotic deterministic randomness system of Eq.(3a) is described as follows:

$$X_{n+1} = \lfloor a \Box X_n \rfloor (\bmod 2^n) \tag{4a}$$

$$Y_n = b \Box X_n (\bmod 2^n) \tag{4b}$$

Furthermore, the discretized model of Eq.(3b) is:

$$X_{n+1} = \begin{cases} \lfloor a \Box X_n \rfloor (\bmod 2^n) & \text{if } i = 0 \\ N - \lfloor a \Box X_n \rfloor (\bmod 2^n) & \text{if } i = 1 \end{cases} \tag{5a}$$

$$Y_{n+1} = \begin{cases} \lfloor b \Box X_n \rfloor (\bmod 2^n) & \text{if } j = 0 \\ N - \lfloor b \Box X_n \rfloor (\bmod 2^n) & \text{if } j = 1 \end{cases} \tag{5b}$$

where $i = \bmod(\lfloor \lfloor aX_n \rfloor / N \rfloor, 2)$. $j = \bmod(\lfloor \lfloor bX_n \rfloor / N \rfloor, 2)$. When $2^n = 2^{51}$, the first return maps of Eq.(4b) and Eq.(5b) are shown in Fig.2.

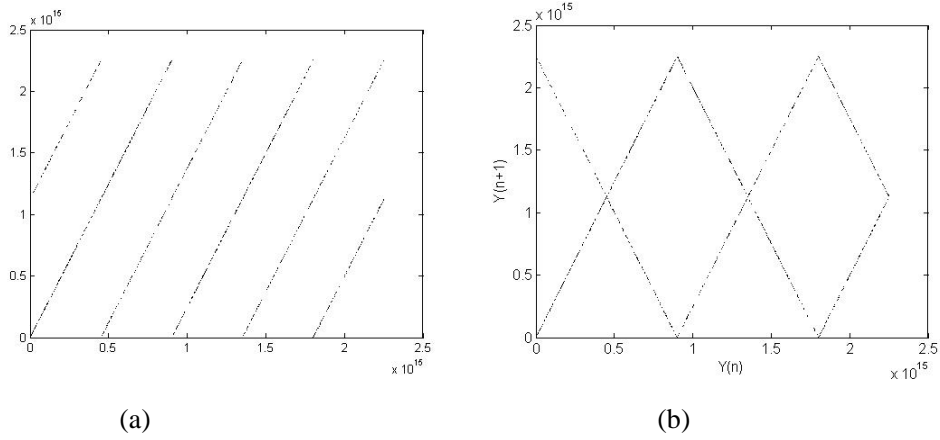

FIG.2: First-return maps produced by Eq.(5a) and Eq.(5b) for $a = 5/2$: (a) $x_{n+1} = h(a, x_n), y_n = h(b, x_n)$; (b) $x_{n+1} = g(a, x_n), y_n = g(b, x_n)$

## III. THE STATISTIC CHARACTERISTICS OF ASYMPTOTIC DETERMINISTIC RANDOMNESS SEQUENCES

Bedsides those characteristics which chaotic maps also have, such as ergodicity, sensitive dependence on initial conditions and random-like behaviors, the asymptotic deterministic randomness sequences have the characteristic of multi-value correspondence, which makes those sequences unpredictable in short steps. That means the asymptotic deterministic randomness sequences have evident advantages in cryptography compared with chaotic sequences. According to **Theorem 1,** the asymptotic deterministic randomness sequence is unpredictable in $N$ steps when $b = q^N$ obviously and will be predictable in $N+1$ steps in almost all the cases. Given the sequence $\{y_n\}_{n=0}^{N-1}$, we can deduce the function $y_{n+N} = g_N(y_n, y_{n+1}, \cdots, y_{n+N-1})$ by enumerating every possible initial values $bf^{-1}(x_n)$, and the computational complexity is $Nq$. When $b \to \infty$, the sequence of the asymptotic deterministic randomness is unpredictable in any steps.

Because Eq.(2a) and Eq.(2b) have the same mechanism, they will have the same dynamical characteristics. Let's take Eq.(2a) as an example. We will analyze the performance of (2a) from a theoretical point of view in the following in order to quantify its statistical properties.

When the control parameter $a \in \Box$, the probability distribution of $\{x_n\}$ is quite uniform. When $a \in \Box$, the probability density of $\{x_n\}$ will converge to a probability density $f_*$ [19]. Let the probability density of $x_n$ be $f_n(x)$, and then the probability density of $x_{n+1}$ is related to that of $x_n$ by

$$f_{n+1}(x) = \sum_{y \in S^{-1}(x)} \frac{f_n(y)}{|S'(y)|} \qquad (6)$$

where Eq.(6) is clearly linear and it is known as the Frobenius-Perron operator. When

$S(x) = ax \bmod 1$, $a \in \mathbb{R}$, the Eq.(6) can be rewritten as:

$$f_{n+1}(x) = \frac{1}{a} \sum_m f_n(ax - m) \tag{7}$$

We will denote the Frobenius-Perron operator with $O$, and then $O^n f_0 = f_n = f_*$. $S(-x) = 1 - S(x)$, and thus it follows that

$$f_{n+1}(-x) = \frac{1}{a} \sum_{y \in S^{-1}(x)} f_n(1 - y)$$

Because $f_*(x)$ and $f_*(-x)$ are all invariant density respectively and there is only one invariant density, it follows that $f_n(y) = f_n(1 - y)$, that is, $f_*$ is even. In Ref.[20], the probability density of another kind of realizable model of asymptotic deterministic randomness, which is named Lissajous map, is given. With the similar means, we will compute $f_*$ by using of Eq.(6,7). Let's Suppose the sequence $\{x_n\}$ have the uniform probability density, i.e. $f_{n+1} = f_n = f_* = 1$. The average number of $m$ is $a$ in interval $[0,1]$ because $m \in \{0, 1, \cdots, \lfloor a \rfloor\}$ when $y > a - \lfloor a \rfloor$ and $m \in \{0, 1, \cdots, \lfloor a \rfloor + 1\}$ when $y \leq a - \lfloor a \rfloor$, The right part of Eq.(7) is equal to 1. Therefore, the sequence $\{x_n\}$ have the uniform probability density in interval $[0,1]$. The probability density of sequence $\{y_n\}$ can also be calculated as uniform easily. The probability distribution of Eq.(3a) with different control parameters is given in Fig.3. It is apparent that the distribution is quite uniform.

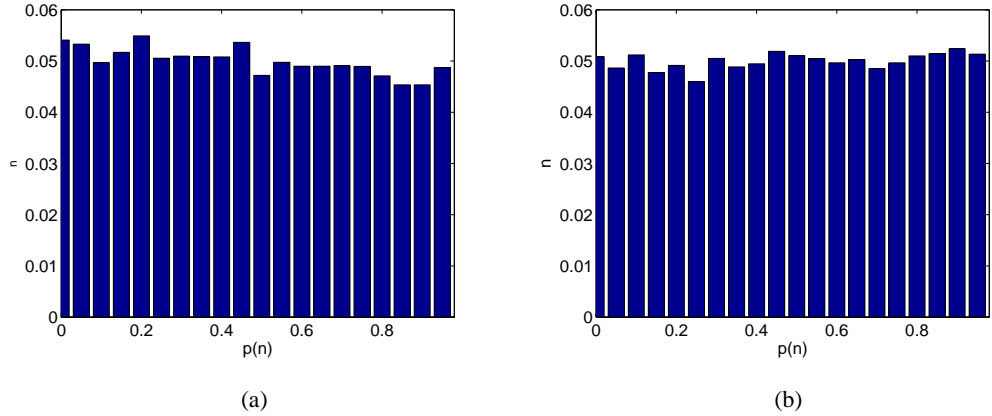

(a)　　　　　　　　　　　　　　　　(b)

FIG.3: The statistical properties for Eq. (3a). (a) $a = 5/2, b = 2$; (b) $a = 7/3, b = 3$

As shown in **Theorem 1**, if we transform the sequence $\{x_n\}$, which is generated by the piecewise linear map $f(x_n) = ax_n \bmod 1$, by using the noninvertible nonlinearity transform $g(x_n) = bx_n \bmod 1$, the sequence $\{y_n\}$ will have the characteristic of multi-value correspondence.

It is known that chaotic sequences have very good random-like properties, and then the sequence $\{y_n\}$ will also have good random-like properties, because the noninvertible nonlinearity transform does not reduce the random-like properties of sequence[18]. In Ref.[21], a new method has been developed, which allows to compare the randomness of deterministic randomness sequences of equal length. Suppose we have a sequence of values $U_1, U_2, \cdots, U_n$. Form a sequence of vectors $X_{(i)} = [U_i, U_{i+1}, \cdots, U_{i+m-1}]$, we can define some variables:

$$C_i^m(r) = \frac{F_j(d[X_{(i)}, X_{(j)}] \leq r)}{N - m + 1}$$

where $d[X_{(i)}, X_{(j)}]$ is the distance between two vectors, which is defined as follows: $d[X_{(i)}, X_{(j)}] = \max(|U_{i+k-1} - U_{j+k-1}|), k = 1, 2, \cdots, m$. The function $F_j(\square)$ returns the total number of j, which makes $d[X_{(i)}, X_{(j)}] \leq r$ hold.

Let's define the possible randomness measurement: $R(m, r, N) = \phi_{(r)}^m - \phi_{(r)}^{m+1}$, where:

$$\phi_{(r)}^m = \frac{1}{N - m + 1} \sum_{i=1}^{N-m+1} \ln C_i^m(r)$$

This method depends on the resolution parameter $r$ and an "embedding" parameter $m$. The calculation of $R$ allows to address the system randomness. This technique has been proved to be very effective in determining system complexity. When $r = 0.025$, $m = 1, 2, 3, 4, 5$ and $N = 10000$, let's calculate the $R(m, r, N)$ of Logistic chaotic sequence, 3D-Chebyshev chaotic sequence, the sequence generated by Lissajous map, and the asymptotic deterministic randomness sequence (A: $a = 5/2$, $b = 2^5$; B: $a = 5/2$, $b = 2^{15}$; C: $a = 11/5$, $b = 2^5$). As shown in Table 1, the two groups of $R(m, r, N)$ of asymptotic deterministic randomness sequences are bigger than any others when $m > 2$. It should be noticed that the randomness of deterministic randomness sequences will be improved if we increase the value of $p$ and $q$, and $R(m, r, N)$ does not change distinctly when control parameter $b$ becomes larger, though the asymptotic deterministic randomness sequence is even harder to predict.

TABLE 1: The possible randomness measurement

|  | $m=1$ | $m=2$ | $m=3$ | $m=4$ | $m=5$ |
|---|---|---|---|---|---|
| Logistic map | -3.5659 | -4.3560 | -5.0501 | -5.7333 | -6.3996 |
| Chebyshev map (n=3) | -3.5495 | -4.6883 | -5.7577 | -6.8027 | -7.7611 |
| piecewise linear map | -3.0114 | -4.1794 | -5.3355 | -6.4557 | -7.4671 |
| Lissajous map | -3.0135 | -4.6209 | -6.1858 | -7.5622 | -8.2663 |
| Asymptotic deterministic randomness A | -3.0122 | -4.6218 | -6.2123 | -7.7154 | -8.7437 |
| Asymptotic deterministic randomness B | -3.0106 | -4.6201 | -6.2117 | -7.7044 | -8.7363 |
| Asymptotic deterministic randomness C | -4.5715 | -6.0131 | -7.6861 | -8.9799 | -9.1870 |

The relation between control parameters $a = p/q$ and the randomness of deterministic randomness has been studied in Refs.[15-16]. Let's study the probability of the deterministic randomness sequence in the following way:

$$P[\alpha_1 < y_n < \alpha_2] = \int_{\alpha_1}^{\alpha_2} f(y_n) dy_n$$

$$P[\alpha_1 < y_n < \alpha_2; \beta_1 < y_{n+1} < \beta_2] = \int_{\alpha_1}^{\alpha_2} \int_{\beta_1}^{\beta_2} f(y_n, y_{n+1}) dy_n y_{n+1}$$

where the function $f(\square)$ is the probability density function or the joint probability density function. If the sequence $\{y_n\}$ and $\{y_{n+1}\}$ are statistically independent, the following equations will be held:

$$P[\alpha_1 < y_n < \alpha_2]\square P[\beta_1 < y_{n+1} < \beta_2] = P[\alpha_1 < y_n < \alpha_2; \beta_1 < y_{n+1} < \beta_2]$$

According to the above conclusion, the probability density functions of the sequences $\{y_n\}$ and $\{y_{n+1}\}$ are 1. However, when the value of $y_n$ is known, the value of $y_{n+1}$ will have only $q$ different values based on the multi-value correspondence, which means the vector $(y_n, y_{n+1})$ is not ergodic in $[-1,1] \times [-1,1]$ and the joint probability density function $f(y_n, y_{n+1})$ can not be 1. The joint probability density function $f(y_n, y_{n+1})$ is shown in Fig.4 when the control parameters $a = 5/2$ and $b = 2^5$. We can see the sequence $\{y_n\}$ and $\{y_{n+1}\}$ are not statistically independent. When we increase the values of $p$ and $q$, the multi-value correspondence will become more complex. If $p, q \to \infty$ (When the control parameter $a$ is an irrational number, we can think $p, q \to \infty$), it means that the value of $y_{n+1}$ can be selected as any values and the vector $(y_n, y_{n+1})$ is ergodic in $[-1,1] \times [-1,1]$. As shown in Fig.5, the vector $(y_n, y_{n+1})$ is ergodic in $[-1,1] \times [-1,1]$ and the joint probability density function $f(y_n, y_{n+1})$ is 1, when $p, q \to \infty$. As a result, if we increase the value of $p$ and $q$, the statistical independence and the randomness of the sequence $\{y_n\}$ will be improved.

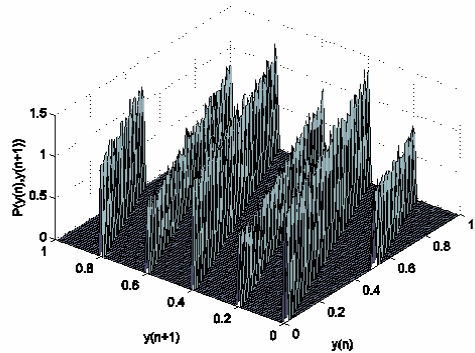

FIG.4: The probability density function when $a = 5/2$

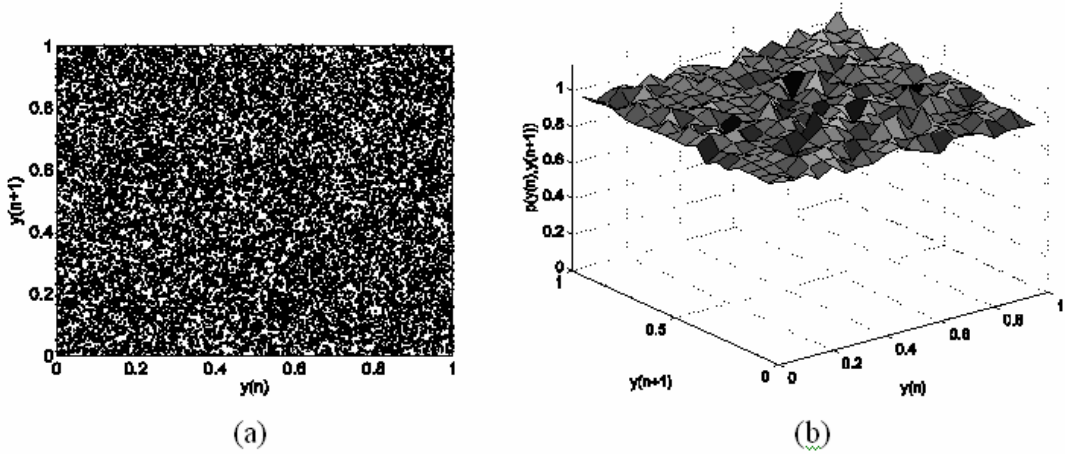

FIG. 5: When $p, q \to \infty$, (a) the first return map, (b) the probability density function

When we increase the value of $b$, the first return map of the system does not change and the ergodicity of vector $(y_n, y_{n+1})$ does not change, either. As a result, the control parameter $b$ does not have significant influence on the independence of the sequences $\{y_n\}$ and $\{y_{n+1}\}$. In Table 2, we study the 2-order correlation of the sequences with different $b$. As shown in Table 2, the sequences $\{y_n\}$ and $\{y_{n+1}\}$ are not statistically independent because $E(y_n^2)E(y_{n+1}^5) \neq E(y_n^2 \Box y_{n+1}^5)$.

TABLE 2: The 2-order correlation between $\{x_n\}$ and $\{x_{n+1}\}$ with different control parameter $b$ $x_{n+1} = \mod(ax_n, 1), y_n = \mod(bx_n, 1)$

|  | $E(y_n^2)$ | $E(y_{n+1}^5)$ | $E(y_n^2 \Box y_{n+1}^5)$ | $E(y_n^2)E(y_{n+1}^5)$ |
| --- | --- | --- | --- | --- |
| $a = 5/2, b = 2$ | 0.3230 | 0.1602 | 0.0566 | 0.0517 |
| $a = 5/2, b = 2^{10}$ | 0.3332 | 0.1665 | 0.0638 | 0.0555 |
| $a = 5/2, b = 2^{32}$ | 0.3333 | 0.1667 | 0.0639 | 0.0556 |
| $a = 5/2, b = 2^{50}$ | 0.3289 | 0.1621 | 0.0611 | 0.0533 |

In our opinion, the wrong impression, that the independence of the sequence can be improved when we increase the value of $b$, is caused by computational precision. If the value of $b$ is larger than the computational precision, the first return of the system will be changed and the independence of the sequences $\{y_n\}$ and $\{y_{n+1}\}$ is also changed. Let's take Lissajous map $x_{n+1} = \cos(a \cos^{-1}(x_n)), y_n = \cos(b \cos^{-1}(x_n))$, $a = 5/2$ as an example. Because the computational precision of **matlab** is $2^{50}$, the statistical character of the sequence will not change if $b < 2^{50}$. The values of $E(y_n^2), E(y_{n+1}^5), E(y_n^2 \Box y_{n+1}^5)$ and $E(y_n^2)E(y_{n+1}^5)$ are 0.5005, $10^{-4}$, 0.0163 and $10^{-4}$, respectively. However, if $b > 2^{50}$ (for example $b = 3^{50}$), as shown in Fig.6, the first return map of the system will change and the value of $E(y_n^2 \Box y_{n+1}^5)$ will be $10^{-4}$.

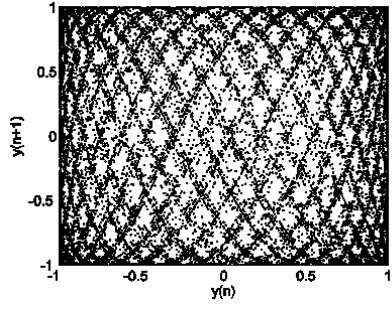

FIG.6: The first return map the system ($a = 5/2$) when the value of $b$ is larger than the computational precision

## IV. THE STATISTIC CHARACTERISTICS OF ASYMPTOTIC DETERMINISTIC RANDOMNESS SYMBOLIC SEQUENCE

In this paper, we generate PRBGs by digitizing the output of asymptotic deterministic randomness. Let's divide the interval $I$ into several subinterval $\beta = \{C_0, \cdots, C_{m-1}\}$ and label them with definite integers which belong to $[0, m-1]$, where $C_i \cap C_j = \phi, i \neq j$ and $\bigcup_{i=0}^{m-1} C_i = I$. When the output $y$ enters the $r$th sub-interval, a pseudorandom number takes an integer $s \in [0, m-1]$. If we generate the pseudorandom number sequence by using a wrong subinterval set $\beta$, the statistic characteristics of the pseudorandom number sequence will tell the attacker some information of the secret key. Let's take the piecewise linear map $f(x_n) = x_n/p, x_n \in [0, p), f(x_n) = (1-x_n)/(1-p), x_n \in [p, 1]$ as an example. If we utilize the subinterval set $\beta = \{[0, 0.5], [0.5, 1]\}$, the probabilities of the sequences 00,01,11 and 10 are $p(s_{n+1}s_n = 01) = 1-p$, $p(s_{n+1}s_n = 11) = p$, $p(s_{n+1}s_n = 00) = p$, $p(s_{n+1}s_n = 10) = 1-p$. When $p = 0.1$, the total number of 00, 01, 10 and 11 are 495, 4489, 4488, 526 respectively in the first 9998 items of the piecewise linear chaotic bit sequence. Therefore, the PRBGs, which utilize the subinterval set $\beta = \{[0, 0.5], [0.5, 1]\}$ can not defend against the entropy attacks.

**Theorem 2:** When $a = p/q > 2$, $b = q^N$, the next value $s_M$ can take $q$ different values equiprobably no matter what the sequence $\{s_n\}_{n=0}^{M-1}$ is, if we utilize the subinterval set $\beta = \{C_i = [(i-1)/q, i/q]\}$, $i = 1, 2, \cdots, q-1$, i.e. $p(s_N = s \mid \{s_n\}_{n=0}^{M-1} \text{ is known}) = \frac{1}{q}$, $s = 1, 2, \cdots, q-1$.

**Proof:** When $s_0 = j$, that means $y_0 \in [\frac{j-1}{q}, \frac{j}{q})$ and $x_0 \in \bigcup_{i=0}^{b-1} I_i, I_i = [\frac{j-1}{qb} + \frac{i}{b}, \frac{j}{qb} + \frac{i}{b})$ according to Eq.(3a). Because the sequence $\{x_n\}$ has the uniform probability density in interval $[0,1]$, $x_0$ is distributed in $I_i$ uniformly and $p(x_0 \in I_i) = \frac{1}{q}, i = 0, 1, \cdots, q-1$, and then $b\frac{p}{q}x_0 \in \bigcup_{i=0}^{b-1}[\frac{p}{q}\frac{j-1}{q} + \frac{p}{q}i, \frac{p}{q}\frac{j}{q} + \frac{p}{q}i)$ will be held. Because $y_1 = \mod(bx_1, 1) = \mod(b\frac{p}{q}x_0, 1)$, $y_1 \in \bigcup_{i=0}^{b-1}[\frac{p}{q}\frac{j-1}{q} + \mod(\frac{p}{q}i, 1), \frac{p}{q}\frac{j}{q} + \mod(\frac{p}{q}i, 1))$. Since $p$, $q$ are relative prime numbers, $p(y_1 \in [\frac{p}{q}\frac{j-1}{q} + \frac{i}{q}, \frac{p}{q}\frac{j}{q} + \frac{i}{q})) = \frac{1}{q}$ will be held. Therefore $p(s_1 = s \mid s_0 = j) = \frac{1}{q}$. Similar deduction can be performed to reach the same conclusion for $p(s_1 = s \mid \{s_0 \text{ is known}\}) = \frac{1}{q}$.

When the sequence $\{s_n\}_{n=0}^{M-1}$ is known, let's denote the intervals where $y_N$ satisfies this condition as $L_i = [\frac{a_i}{c_i}, \frac{b_i}{c_i}]$, $i = 0, 1, \cdots$, where $y_N$ is distributed uniformly in each $I_i$. When $y_N \in L_i$, similar deduction can be performed to reach the following conclusion: $p(y_{N+1} \in [\frac{p}{q}\frac{a_i}{c_i} + \frac{i}{q}, \frac{p}{q}\frac{b_i}{c_i} + \frac{i}{q}) | y_N \in L_i) = \frac{1}{q}$. Therefore $p(s_N = s | \{s_n\}_{n=0}^{N-1} \text{ is known}) = \frac{1}{q}$.

Let's calculate the entropy $H_n^\beta = -\sum_{s^n} P(s^n) \log P(s^n)$ of the symbolic sequence $\{s_n\}$, where $\{s_n\}$ generated by Eq.(3a) and the $P(s^n)$ means the probability of each sequence with $n$ length. When symbolic sequence $\{s_n\}$ satisfies $p(s_1 = s | \{s_0 \text{ is known}\}) = \frac{1}{q}$, entropy and conditional entropy will be $H_n^\beta = n$ and $h_n^\beta = 1$ respectively [1,2]. When $a = 5/2$, $b = 2$ and $\beta = \{[0, 1/2], [1/2, 1]\}$, entropy and conditional entropy of the symbolic sequence $\{s_n\}$ is shown in Fig 7.

$$h_n^\beta = H_{n+1|n}^\beta = \begin{cases} H_{n+1}^\beta - H_n^\beta, & n \geq 1 \\ H_1^\beta, & n = 0 \end{cases}$$

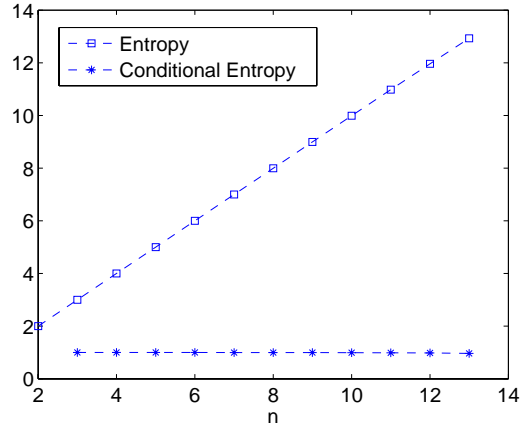

FIG. 7: Entropy and conditional entropy of the symbolic sequence $\{s_n\}$

The auto-correlation and cross-correlation of the symbolic sequence $\{s_n\}$ are also given in Fig8(a,b). It can be seen that $\{s_n\}$ has $\delta$-like auto-correlation which is required for a good PRBGs. The sequences generated with different initial values will have zero cross-correlation due to the sensitive dependence on initial conditions.

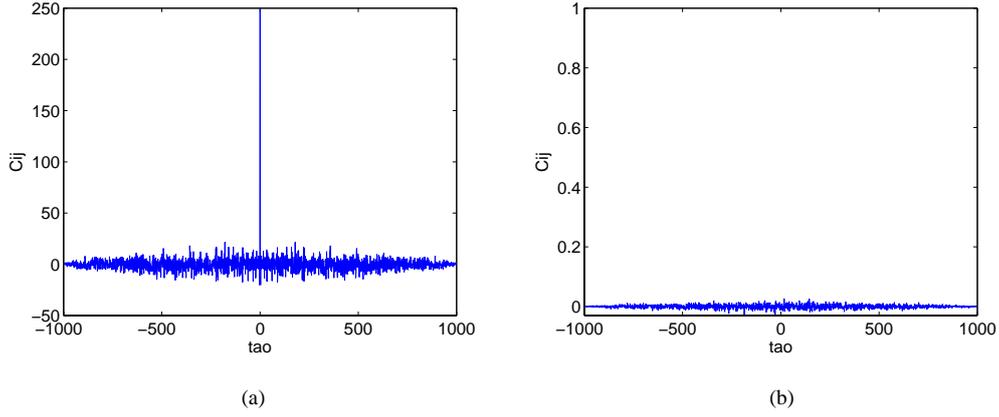

FIG.8: The auto-correlation and cross-correlation of the symbolic sequence $\{s_n\}$

When $s_n$ is known, the inverse chaotic map $x_n = f_{s_n}^{-1}(x_{n+1})$ will be confirmed unambiguously. Any value selected in interval $I$ will converge to its initial condition with respect to sufficient backward iterations [22]. Because of those properties, the chaotic PRBGs can not resist the cryptanalysis strategy using symbolic dynamics. As a result, many symmetric and asymmetric encryptions, which use pseudorandom bit sequences generated by chaotic systems, have been proved to be insecure. Similarly, the multiple PRBGs based on the spatiotemporal chaotic map are also proved be insecure, because they can not resist the cryptanalysis strategy based on symbolic vector dynamics.

The key of the cryptanalysis strategy using symbolic dynamics is to calculate the inverse chaotic function. Because of the characteristic of multi-value correspondence, we can not calculate the inverse function of asymptotic deterministic randomness when the symbolic sequence $\{s_n\}$ is known, which improves the security of chaos based PRBGs and increases the resistance against symbolic dynamics attacks.

We will explore the relation between the symbolic sequence of the system and the initial value by using the Gray Ordering Number: $G_N(S) = \sum_{n=0}^{N} g_n 2^{-(n+1)}$, where $g_{n+1} = s_{n+1} + g_n (\mod 2)$, $g_0 = s_0$ and $x = 0.s_0 s_1 \cdots s_N$. From Fig.9(a,b), we can see that $G_N(S)$ of Logistic map and skewed tent map are monotonic increasing functions with the increasing of initial values, and we can deduce the initial values according to the symbolic sequence without difficulty. Unlike the chaotic map, the same initial value $y_0$ of the deterministic randomness may correspond to any symbolic sequence with the equal probability due to the characteristic of multi-value correspondence. The symbolic dynamics is more complex when the control parameter $b$ becomes larger.

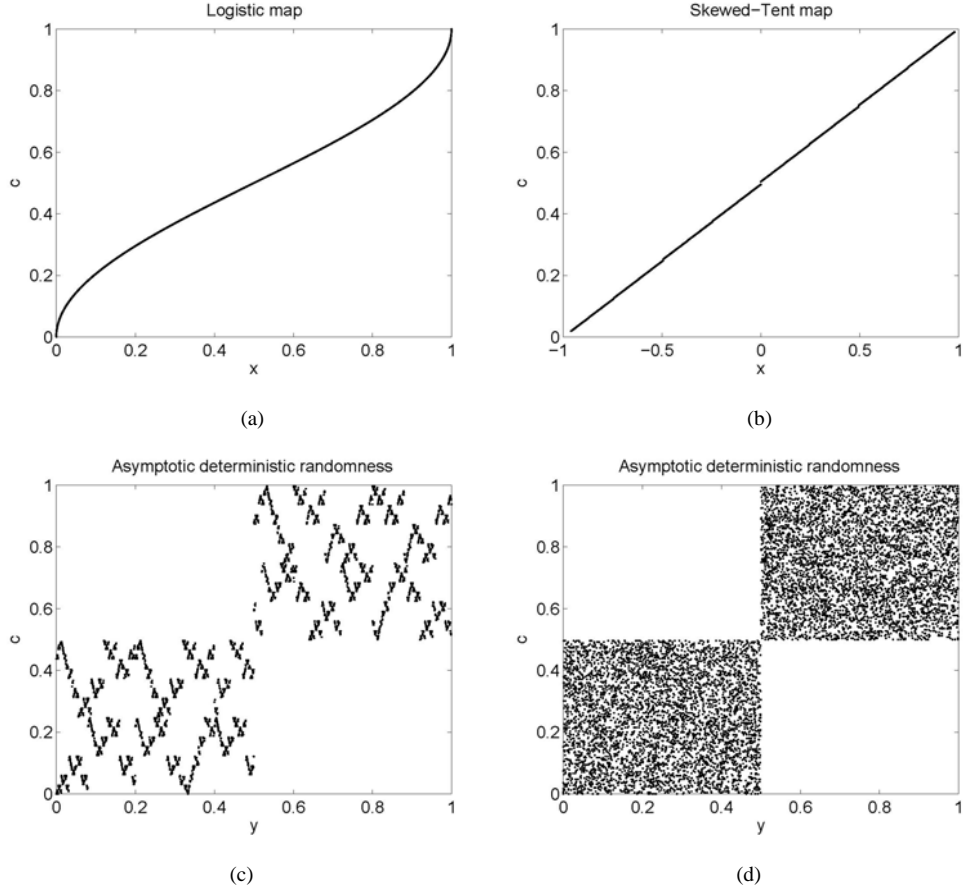

FIG. 9: The relation between the symbolic sequence and the initial value, $N=8$: (a) Logistic map: $F(x_n)=4x_n(1-x_n)$; (b) The Skewed-Tent map: $f_i(x)=p-1-p|x|, p=1.99$; (c) Asymptotic deterministic randomness, $b=2$; (d) Asymptotic deterministic randomness, $b=2^{20}$

## V. CONCLUSION

In this paper, we study the characteristic of multi-value correspondence of the asymptotic deterministic randomness, which is constructed by the piecewise linear map and the noninvertible nonlinearity transform; we also give the discretized systems in the finite digitized state space. The statistic characteristics, such as stationary probability density function and random-like behavior, are investigated numerically. The probability density of asymptotic deterministic randomness sequence $\{y_n\}$ can be calculated as uniform in interval $[0,1]$. Furthermore, the symbolic dynamics of asymptotic deterministic randomness sequence are also investigated. When $a=p/q>2$, $b=q^N$, the next value $s_M$ can take $q$ different values equiprobably no matter what the sequence $\{s_n\}_{n=0}^{M-1}$ is, if we utilize the subinterval set $\beta=\{C_i=[(i-1)/q,i/q]\}$, $i=1,2,\cdots,q-1$. The symbolic sequence $\{s_n\}$ has $\delta$-like auto-correlation, zero cross-correlation and the complex relation between the symbolic sequence and the initial value. As a result, the symbolic sequence of the asymptotic deterministic randomness possesses very good cryptographic properties, which increase the resistance against entropy attacks and the attacks using symbolic dynamics.

## ACKNOWLEDGEMENTS

This work was supported by the Natural Science Foundation of China under Grant 60672095, the National High Technology Project of China under Grant 2002AA143010 and 2003AA143040, the Excellent Young Teachers Program of Southeast University.